\documentclass[%prl,
preprint,
 amssymb,floatfix,amsmath,aps,superscriptaddress,showpacs]{revtex4}
\usepackage{graphicx}
\usepackage{dcolumn}
\usepackage{bm}

\begin{document}

\title{Observation of two species of vortices in the anisotropic\\ spin-triplet superconductor Sr$_{2}$RuO$_{4}$}

\author{V.O. Dolocan \footnote{Present address: Max Plank Institute for Chemical Physics of Solids, 40 N\"{o}thnitzer
Str., 01187 Dresden, Germany}}\affiliation{CRTBT-CNRS, 25 Avenue des
Martyrs, 38042 Grenoble, France}
\author{P. Lejay}\affiliation{CRTBT-CNRS, 25 Avenue des Martyrs, 38042 Grenoble, France}
\author{D. Mailly}\affiliation{LPN-CNRS, Route de Nozay, 91460 Marcoussis, France}
\author{K. Hasselbach}\affiliation{CRTBT-CNRS, 25 Avenue des Martyrs, 38042 Grenoble, France}
\email{Second.Author@institution.edu}

\begin{abstract}
Magnetic flux structures in single crystals of the layered spin
triplet superconductor Sr$_{2}$RuO$_{4}$ are studied by scanning micro SQUID Force microscopy. Vortex chains
appear as the applied field is tilted along the in-plane direction of the superconductor. The vortex chains align along the direction of the in-plane component of the applied magnetic field.
The decoration of in-plane vortices by crossing Abrikosov vortices is observed: two vortex orientations are apparent simultaneously, one along the layers and the other perpendicular to the layers. The crossing vortices appear preferentially on the in-plane vortices.
\end{abstract}

\date{\today}
 \maketitle

\section{Introduction}
The magnetic properties of superconductors depend strongly on their crystalline and electronic anisotropy. Prominent examples are the high temperature superconductors. Intense research efforts have revealed a variety of vortex phenomena in these compounds (vortex-glass, vortex melting, crossing of Josephson and pancake vortices). The general theoretical approach on vortex matter is based on the anisotropic Ginzburg-Landau (GL) theory. There the
anisotropy is expressed in terms of the effective mass of the electron. For layered anisotropic superconductors, the out of plane effective mass, m$_{c}$, is much larger than the in plane effective masses (m$_{c}$$>>$m$_{ab}$). To describe this anisotropy the parameter
$\gamma$=(m$_{c}$/m$_{ab}$)$^{1/2}$=$\lambda_{c}/\lambda_{ab}$\cite{Clem}
is used. For example in NbSe$_{2}$ $\gamma$=3.3, in YBCO$\gamma$=5-8
and in BSCCO $\gamma$ is higher than 150, $\gamma$ being dependent on the oxygen doping of the high T$_{c}$ superconductors.

In moderately anisotropic superconductors, when the magnetic field
is tilted away from the anisotropy axis, calculations based on GL
theory show that the screening currents tend to flow in approximately
elliptical paths around the vortex cores and parallel to the
layers\cite{Buzdin,Kogan}. This current flow creates a net
transverse magnetization and attraction appears between the tilted
vortices, leading finally to the development of vortex chains,
formed by inclined vortices. These vortex chains have been observed
in YBCO\cite{Gammel}.

In highly anisotropic superconductors, Josephson vortices appear if
the distance between the layers is larger than $\xi_{c}$. The
Josephson vortices are confined in the space between the layers. The
Lawrence-Doniach (LD) model is used as the continuous GL theory
cannot be applied anymore. In the LD model the tilted vortex is
described as tilted stack of 2D pancake vortices connected by
Josephson strings. Pancake vortices are characterized by circular
supercurrent pattern flowing in the layers and Josephson strings
being short segments of Josephson vortices confined to the
insulating region between the superconducting layers. High T$_{c}$
superconductors with a weak interlayer coupling, like BSCCO, are
treated as 2D superconductors according to the Lawrence-Doniach
model. Magnetic imaging of vortices was quintessential for the
experimental exploration of vortex structures in BSCCO: Bitter
decoration experiments revealed vortex chains separated by a vortex
lattice\cite{Bolle}, scanning Hall probe microscopy showed Josephson
vortices decorated by pancake vortices\cite{Grigorenko, Bending_2005}.

In the present study vortex matter is investigated in the anisotropic layered
superconductor Sr$_{2}$RuO$_{4}$. The
superconducting transition temperature of Sr$_{2}$RuO$_{4}$ is of
the order of 1.45 K\cite{Maeno}. Like high T$_{c}$ cuprates, the
tetragonal Sr$_{2}$RuO$_{4}$ has a layered structure, the RuO$_{2}$
planes are separated by 12.74 {\AA} and has highly anisotropic
properties\cite{MM}. Sr$_{2}$RuO$_{4}$ has a $\gamma$ value of 20
situating it between YBCO and BSCCO on the anisotropy scale. We
expect Sr$_{2}$RuO$_{4}$ to act more like a 3D superconductor as the
c-axis parameter is 3 times smaller than the coherence length
$\xi_{c}$. The Ginzburg-Landau parameter $\kappa$ = $\lambda/\xi$ is
around 2.3 when the magnetic field is applied along the c-axis direction and 46 for the
in-plane direction. The physical properties of
Sr$_{2}$RuO$_{4}$ are very rich and indications accumulated that it
might exhibit unconventional
superconductivity\cite{Ishida,Ying,Luke}. A coherent picture can be
obtained in terms of a superconducting state characterized by
breaking of time reversal symmetry and requiring a multi-component
order parameter of p-wave symmetry with a gap function
d=$\widehat{z}$(k$_{x}\pm$ik$_{y}$). The symmetry of the order
parameter should give raise to domains of opposite chirality,
$\widehat{z}$(k$_{x}$+ik$_{y}$) and $\widehat{z}$(k$_{x}$-ik$_{y}$),
separated by domain walls, formation of fractional vortices on the
domain walls and should generate spontaneous currents at the edges
of the crystal. It was conjectured that domain walls might be preferential pinning sites for vortices\cite{Sigrist}. Magnetic
microscopy is a means of choice to study the relevance of these
predictions. Vortices along with vortex coalescence were observed in
Sr$_{2}$RuO$_{4}$ by $\mu$SQUID magnetic microscopy \cite{Dolocan}.
Here we focus on vortex matter in Sr$_{2}$RuO$_{4}$, demonstrating
the presence of tilted vortex chains in, and evidencing pinning
of perpendicular vortices on in-plane vortex chains in
Sr$_{2}$RuO$_{4}$, making Sr$_{2}$RuO$_{4}$ the first anisotropic 3D
superconductor presenting crossing vortices.

We use for magnetic imaging a high resolution scanning $\mu$SQUID
microscope (S$\mu$SM)\cite{Veauvy} working in a dilution
refrigerator. The S$\mu$SM has an aluminum $\mu$SQUID as pickup loop
of 1.2$\mu$m diameter. The critical current of the $\mu$SQUID is a
periodic function of the magnetic flux emerging perpendicularly from the sample surface. The images shown are maps of the critical current value of the $\mu$SQUID scanning the surface. We used two different high quality crystals coming from the same rod showing volume superconductivity below a temperature of 1.31K. The two
samples are large plates with thickness of 0.5mm and 0.6mm. The
surface is cleaved and AFM images show flatness down to the order of
6 $\AA$. The magnetic fields are applied by a solenoid and a
rotatable Helmholtz coil, the coils are at room temperature. The solenoid axis is
parallel to the ab face of the sample (H$_{a}$ in the
Fig.~\ref{Fig.1}e) and the Helmholtz coil generates a field,
(H$_{bc}$), perpendicular to the solenoid axis. Adjusting the relative angle and the magnitude of the two fields allow us to point the resultant field along any direction.

\section{Vortex Chains in Sr$_{2}$RuO$_{4}$}
\begin{figure}
\includegraphics[width=8cm]{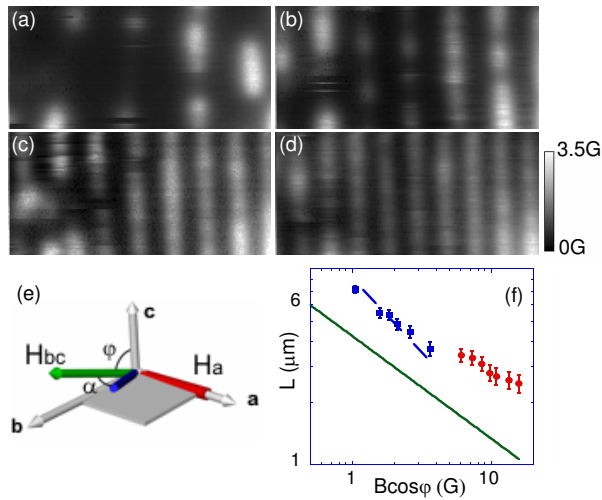}\\
 \caption{\label{Fig.1}(Color online) Scanning $\mu$SQUID microscope images of
 the magnetic flux above the ab-face of Sr$_{2}$RuO$_{4}$ at T=0.35K.
 During field cooling the magnetic field is applied at an fixed angle
 $\varphi$=87$^{\circ}$ from c-axis varying for each panel as: (a)20 G,
 (b) 35 G, (c) 60 G and (d) 70 G. The image area is 31$\mu$m$\times$15$\mu$m.
 The magnetic scale is shown on the right. (e)Scheme of the 2 applied magnetic fields.
 The magnetic field H$_{bc}$ can be rotated with an angle $\varphi$ from the c-axis (in the bc-plane).
 A second fixed magnetic field H$_{a}$ can be applied in-plane. The resultant field is also shown schematically.
 (f)Inter-chain distance as function of normal field. The points are fitted with a B$^{-1}$ law. The line represents
 the isotropic limit (B$^{-1/2}$).}
\end{figure}

Vortex chains form in Sr$_{2}$RuO$_{4}$ \cite{Dolocan2} at low fields and tilt angles higher than 70$^{\circ}$ from c-axis. In
Fig.~\ref{Fig.1} only a field H$_{bc}$ was applied at an angle
$\varphi$ of 87$^{\circ}$ from c axis and the magnetic flux was
increased from 0 to 70 gauss. Each measurement was acquired after a field cooling procedure. The vortex chains start to appear in our images for fields higher than 20 gauss, 2H$_{c1}$ for the in-plane direction of our sample. As the field is increased the chain density
increases also. When the number of vortices reaches an equilibrium value in each chain the formation of a new chain becomes energetically favorable.

Calculations in the GL theory in the London limit\cite{Daemen1} show
that the intra-chain distance between vortices remains constant with
increasing field while the inter-chain distance decreases like B$^{-1}$. This distance can be calculated in the single chain limit
and, using the parameters of Sr$_{2}$RuO$_{4}$, the vortex vortex intra-chain distance yields 2$\lambda_{ab}$$\sim$0.3$\mu$m, lower than our SQUID resolution. On the other hand we can measure the
inter-chain distance between the chains and compare it with the GL
theory. In our experiment on Sr$_{2}$RuO$_{4}$, at low perpendicular fields the distance between the chains follows the B$^{-1}$
dependence as in the GL theory (Fig.~\ref{Fig.1}f, fitted points),
though the value of the observed spacing is larger than predicted.
As the field increases the distance between the chains should tend
monotonously to the isotropic limit (the line, B$^{-1/2}$
dependence). A regular square lattice is observed by SANS for fields higher
than 50G applied along the c-axis. With increasing perpendicular field, we
observe that the inter-chain distance deviates from the B$^{-1}$
dependence characteristic of vortex chains (Fig.~\ref{Fig.1}f,
points not fitted). The chains seem to accommodate a higher density
of vortices than predicted by the London theory, retarding the formation of the vortex lattice. This finding is consistent with
vortex coalescence \cite{Dolocan} in Sr$_{2}$RuO$_{4}$ at
intermediate magnetic fields applied parallel to c-axis.

\begin{figure}[!t]
\includegraphics[width=8cm]{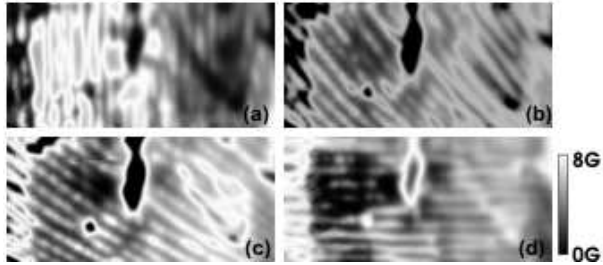}
 \caption{\label{Fig.2}Rotation of vortex chains in Sr$_{2}$RuO$_{4}$.
 When the amplitudes of the two different magnetic fields are varied the angle $\alpha$
 of the resultant field is changed:(a)H$_{bc}$=54G, $\varphi$=80$^{\circ}$, $\alpha$=0$^{\circ}$;
 (b)H$_{bc}$=54G, H$_{a}$=54G, $\varphi$=85$^{\circ}$, $\alpha$=45$^{\circ}$; (c)H$_{bc}$=39.3G,
 H$_{a}$=68G, $\varphi$=85$^{\circ}$, $\alpha$=60$^{\circ}$; (d) H$_{bc}$=-6G, H$_{a}$=90G,
 $\varphi$=0$^{\circ}$, $\alpha$=90$^{\circ}$. All images are taken after field cooling at a temperature of 0.6K. The dimensions of each image are 62$\mu$m$\times$30$\mu$m.} \end{figure}

Ginzburg-Landau theory predicts that the vortex chains are
aligned in the plane spanned by the crystal's anisotropy axis and the applied field. When the applied field is rotated in the ab-plane, the vortex chains should follow. We undertook these measurements on the second crystal, in turning the in-plane component of the field vector  by an angle $\alpha$ relative to the bc-plane. The field was applied in superposing two fields: H$_{a}$ oriented along a-axis, and the field in the bc-plane H$_{bc}$ tilted from the c-axis by the angle $\varphi$.
Each measurement was done after field
cooling (Fig.~\ref{Fig.2}). We varied the direction $\alpha$ of the in-plane component from 0$^{\circ}$ to 90
$^{\circ}$. The linear vortex chains align with the in-plane direction of the applied field as expected. The features in the magnetic images that are fixed are due to flux pinned at surface asperities. In
the panel a) the broadened flux lines are attributed to the
simultaneous presence of vortex chains and vortex coalescence,
fostered by a perpendicular field of 9 G. For some values of the
perpendicular field the vortices in chains are well separated like
in the image c) when the perpendicular field is of the order of 3.5
G. These single vortices seem to be pinned on the vortex chains,
reminiscent of decoration of Josephson vortices by pancake vortices
observed in the 2D superconductor BSCCO. Exploring the history
dependence of the magnetic field penetration in the crystal allows
us to examine this novel state in the anisotropic 3D superconductor in more
detail.

\section{Crossing Vortices in Sr$_{2}$RuO$_{4}$}
\begin{figure}[!]
 \includegraphics[width=8cm]{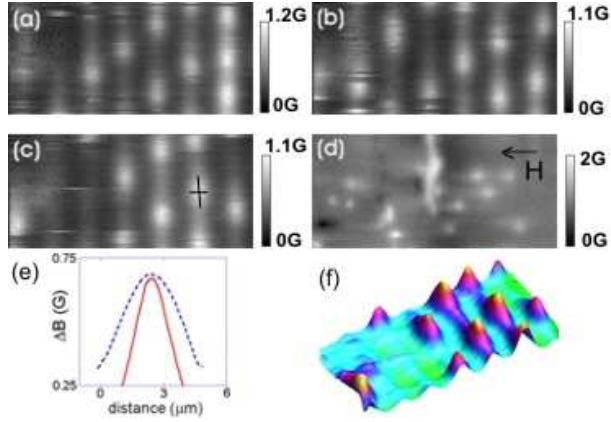}\\
 \caption{\label{Fig.3}(Color online) Crossing vortices in Sr$_{2}$RuO$_{4}$.
 The (a)-(c) images are taken after different field preparations:
 (a)50 G ZFC, (b)5 G FC and increased to 50 G at low temperature, (c) 10 G FC and increased to 50 G at low T. The magnetic scale for each image is shown on the right. The dimensions are 31$\mu$m$\times$15$\mu$m and the imaging temperature is 0.35K. The tilting angle is 87$^{\circ}$. (d)68G FC with field applied only in plane (indicated by an arrow). An 'halo' around each perpendicular vortex is visible. The dimensions of the image are 62$\mu$m$\times$30$\mu$m. (e) Twp perpendicular line scans at one of the decorating vortices (indicated by a cross in panel c). The  image (f), is the 3D representation of the image (c).}
 \end{figure}

In order to explore the vortex pinning and the vortex chain
mechanism in Sr$_{2}$RuO$_{4}$ we undertook measurements changing
the applied magnetic field while the sample is in the
superconducting state. The magnetic field component parallel to the
plane is higher than the first critical field H$_{c1}$  in the in-plane direction (H$_{c1}^{ab}$=10G), while the perpendicular
component is lower than H$_{c1}$(H$_{c1}^{c}$=50G). In a first
approximation a formation of tilted vortex chains might be expected as shown in the field cooled case (see Fig.~\ref{Fig.1}). With
magnetic fields close to the ab-plane for different field
preparations we observe vortices decorating flux channels. We observe also a tendency for a long-range anticorrelation between decorating vortices in adjacent chains reminding a distorted hexagonal lattice. This anticorrelation may be a sign that the vortex
pinning on the chains is weak. Zero field cooling (ZFC)
Fig.~\ref{Fig.3}(a)-(c) and field cooling (FC) Fig.~\ref{Fig.1}c)
sample preparation result in different vortex configurations: we
observe the decoration of flux channels mostly in the ZFC
experiments. Due to the lower first critical field the magnetic flux
penetrates preferentially along the planes, forming the flux channels, and the perpendicular
vortices are pinned on the flux channels. In the FC
experiments for a wide range of angles the flux expulsion happens in such a way that only tilted vortices appear in the sample. In
Fig.~\ref{Fig.2}(c) aligned vortices can be seen. The distinction of
chains of tilted vortices and flux channels along the planes is made
considering the amplitude of the signal. As shown by the magnetic
scales on the right of each image in Fig.~\ref{Fig.3}, in the
(a)-(c) images the vortices have a similar magnetic amplitude as the
vortices present when the field is applied along the
c-axis\cite{Dolocan}. A normal component of the field of the order
of 0.4 Gauss creates these vortices. In the FC case
(Fig.~\ref{Fig.1}c) the magnetic flux coming out of the sample has much larger amplitude as if the vortices would be closer together.
We should have around 70 vortices in the image for the FC case but
the vortices appear as tilted vortices forming a vortex chain. All
the images in Fig.~\ref{Fig.1} and Fig.~\ref{Fig.3}(a)-(c) are taken at the same place of the sample. We observe also that the decorating
vortices have oval shape (Fig.~\ref{Fig.3}d), similar to the
pancake vortices decorating Josephson vortices\cite{Grig05}. In
the case of BSCCO the deformation is explained by a displacement of
the pancake vortices due to circulating currents of the Josephson
vortices\cite{Koshelev}. In a similar manner the interaction between
in-plane flux channels and the crossing vortices  may result in the
elongated shape of the decorating vortices detected by the SQUID.
The perpendicular vortices appear due to a small residual field
estimated to be 0.15G. The deformation of the vortex shape is a sign
for the strong interaction between the two species of vortices and
confirms the finding that the crossing vortices are pinned on the
flux channels running along the layers.

 \begin{figure}
\includegraphics[width=6cm]{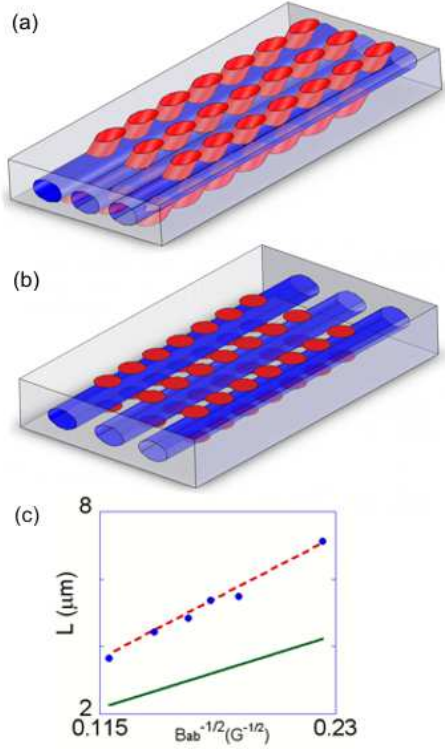}\\
 \caption{\label{Fig.4}(Color online) Scheme of interpenetrating vortices: (a)tilted vortices and
 (b)pancake vortices crossing Josephson vortices. (c)the inter-chain distance as function
  of the in plane component of the field B$_{ab}^{-1/2}$, supposing the applicability of the Lawrence Doniach model.
   The dashed line represents the theoretical values of the inter-chain distance calculated from eq.1 with $\gamma$=28.
   The line represents the values calculated with $\gamma$=20.}
\end{figure}

\section{Discussion}
In isotropic superconductors at low tilted fields the flux lines
penetrate parallel to each other and with the average field in the
sample and  arrange in a pattern that minimizes the interaction
energy. If the superconductor is strongly anisotropic the free
energy of the vortex state has two local minima and consequently a
flux line can penetrate in two distinct directions in the material.
For high anisotropy only flux lines that are nearly parallel or perpendicular to the layers are stable\cite{Huse,Sudbo}. In the LD
model this corresponds to the separation in pancake vortices and
Josephson vortices. The distance between the Josephson vortices is
given by\cite{Koshelev} L=$\sqrt{\sqrt{3}\gamma\phi_{0}/(2B_{ab})}$
(1), where B$_{ab}$ is the in plane field (parallel field) and
$\phi_{0}$ is the quantum of flux. In Sr$_{2}$RuO$_{4}$ this
distance should correspond to the inter-chain distance supposing the
LD model applies. In the Fig~\ref{Fig.4}c) the inter-chain distance
is shown function of B$_{ab}^{-1/2}$. The dashed line passing
through the measured points represents the theoretical values
calculated with the formula (1) for an anisotropy parameter
$\gamma$=28. The line represents the theoretical calculation for a
$\gamma$=20, which corresponds to the values found in the
literature\cite{MM}. The anisotropy parameter $\gamma$ is determined from the ratio of the critical fields or penetration depths. A value
of 15-20 was found also by $\mu$SQUID microscopy, imaging the ac
face of the crystal and determining the deformation of the vortex
(images not shown). As the superconductivity is thought to be
orbital dependent\cite{Agterberg}, the anisotropy parameter could
vary (giving rise to two different penetration depths) and an
anisotropy parameter value of 28 might be plausible. The structures
observed would then correspond to pancake vortices pinned on Josephson vortices
(Fig.~\ref{Fig.4}b). However, Sr$_{2}$RuO$_{4}$ should act more like
a 3D superconductor as the coherence length is three times longer than the interlayer spacing. This makes the observed decoration of
in-plane flux channels by crossing vortices unique.

The presence of Josephson vortices is not a necessary prerequisite
for the presence of crossing vortices: in the classical London
theory two parallel vortices repel each other. This repulsion is
maintained if one tilts one of the vortices from the common
direction. The interaction energy is given by:
E$_{int}=\frac{\phi_{0}^{2}}{2\mu_{0}\lambda_{ab}}\mathrm{cot}(\alpha)
\mathrm{exp}(-\frac{d}{\lambda_{ab}})$ where d is the smallest
distance between vortices and $\alpha$ the angle between
them\cite{Sudbo2}. When the angle between vortices becomes
90$^{\circ}$ the flux lines do not interact. If the angle increases
more the flux lines start to attract each other and the highest attraction energy is for the antiparallel direction. Daemen \textit{et al.}\cite{Daemen2} showed that the free energy of a mixed lattice formed by vortices parallel to the c axis and vortices
inclined at some angle from the c axis is lower than the free energy
of a deformed inclined lattice. This model predicted that two types of vortices may be observed in Sr$_{2}$RuO$_{4}$, formed by two interpenetrating lattices (Fig.~\ref{Fig.4}a), a lattice of vortices parallel to the layers and decorating vortices that are nearly parallel to the c-axis.
\section{Conclusion}
We imaged two types of vortex configurations in the anisotropic spin triplet superconductor Sr$_{2}$RuO$_{4}$. The first configuration are vortex chains formed by tilted vortices. The qualitative variation of the vortex chain spacing is consistent with GL theory though the vortex spacing is closer than predicted. For higher perpendicular fields we see that the variation of the vortex chain spacing is slower than predicted, resulting in a higher vortex density in the chain. This accommodation of a higher vortex density may be related to the observed vortex coalescence in Sr$_{2}$RuO$_{4}$.

The second vortex configuration consists of in-plane flux channels crossed
by perpendicular Abrikosov vortices. The observed deformation of the pinned Abrikosov vortices is a sign for the interaction between the Abrikosov vortices and the flux channels. The strong interlayer coupling
of Sr$_{2}$RuO$_{4}$ excludes the existence of Josephson vortices in
this compound, precluding the applicability of the Lawrence Doniach
model of an interplay between Josephson vortices and pancake
vortices, well established for crossing vortices observed in high
T$_{c}$ superconductors. The possibility of crossing vortices is
predicted in the case of sufficiently anisotropic 3D
superconductors, we think that this is the case here. Our
observations are qualitatively understood, for a more quantitative
interpretation the theoretical models have yet to be developed
taking into account the physical properties of Sr$_{2}$RuO$_{4}$,
quite different from those of other anisotropic superconductors. We
are shedding light on %new physical
aspects of vortex physics %in general
 in exploring vortex chain formation and of flux channels
decorated by vortices and in particular we show the richness of
physical phenomena in Sr$_{2}$RuO$_{4}$. Sr$_{2}$RuO$_{4}$ is at the
intersection of many different domains: unconventional
superconductivity, anisotropy and nonlocal electrodynamics of low
$\kappa$ superconductors.

\begin{acknowledgments}
We acknowledge the support of CNRS, and fruitful discussions with Y.
Liu, P. Rodi\`{e}re and T. Prouv\'{e}.
\end{acknowledgments}

\end{document}